\documentclass[aps,pre,twocolumn,superscriptaddress,longbibliography]{revtex4-1}\usepackage{mathrsfs} 
\usepackage{amssymb} 
\usepackage{graphicx,color} 
\usepackage{bbm} 
\usepackage{multirow}
\usepackage{amsmath}
\usepackage{bbm} 
\usepackage{mathrsfs} 
\usepackage{hyperref}
\usepackage[usenames,dvipsnames]{xcolor}
\usepackage{color}
\definecolor{darkgreen}{rgb}{0,0.6,0}
\definecolor{darkblue}{rgb}{0,0,0.6}
\definecolor{darkred}{rgb}{1,0,0}
\definecolor{darkpurple}{rgb}{0.5,0,0.5}
\hypersetup{
bookmarksopen=true,
pdftitle="",
pdfauthor="", 
pdftoolbar=false, 
pdfstartview={FitH},		
pdfmenubar=true,			
pdfhighlight=/O,			
colorlinks=true,			
urlcolor=darkblue,
citecolor=darkblue,		
linkcolor=darkblue}
\newcommand{\beq}{\begin{equation}} \newcommand{\eeq}{\end{equation}}
\newcommand{\bal}{\begin{align}} \newcommand{\eal}{\end{align}}
\newcommand{\wh}{\widehat}

\newcommand{\argc}[1]{\left[#1\right]}

\def\FF{{\cal F}}
\renewcommand{\phi}{\varphi}

\newcommand{\del}{\Delta}

\def\redv{\bar v}
\let\D=\Delta  \let\io=\infty \let\f=\varphi \let\b=\beta
\newcommand{\afunc}[1]{\operatorname{\mathsf{#1}}} \def\DE{\afunc{D}}

\def\dd{\mathrm d}

\begin{document}

\title{Marginally stable phases in mean-field structural glasses}

\author{Camille Scalliet}

\author{Ludovic Berthier}

\affiliation{Laboratoire Charles Coulomb (L2C), Universit\'e de Montpellier, CNRS, 34095 Montpellier, France}

\author{Francesco Zamponi}
\affiliation{Laboratoire de physique th\'eorique, D\'epartement de physique de l'ENS, 
\'Ecole normale sup\'erieure, PSL Research University, Sorbonne Universit\'es, UPMC Univ. Paris 06, CNRS, 75005 Paris, France}

\date{\today}

\begin{abstract}
A novel form of amorphous matter characterized by marginal stability was recently discovered in the mean-field theory of structural glasses. 
Using this approach, we provide complete phase diagrams delimiting the location of the marginally stable glass phase for a large variety of pair interactions and physical conditions, extensively exploring physical regimes relevant to granular matter, foams, emulsions, hard and soft colloids, and molecular glasses. We find that all types of glasses may become marginally stable, but the extent of the marginally stable phase highly depends on the preparation protocol. 
Our results suggest that marginal phases should be observable for colloidal and non-Brownian particles near jamming, and poorly annealed glasses. For well-annealed glasses, two distinct marginal phases are predicted.
Our study unifies previous results on marginal stability in mean-field models, and will be useful to guide numerical simulations and experiments aimed at detecting marginal stability in finite dimensional amorphous materials. 
\end{abstract}

\maketitle

\section{Introduction} 

Twenty years ago, a unified phase diagram for amorphous matter~\cite{LN98} motivated the  search for similarities and differences between the properties of a broad range of materials, from granular materials to molecular glasses~\cite{LN01}. It is now well-established that in the presence of thermal fluctuations, dense assemblies of atoms, molecules, polymers, colloidal particles undergo a glass transition~\cite{Ca09,BB11} as the temperature is decreased or the density increased. In the absence of thermal fluctuations, solidity instead emerges by compressing particles across the jamming transition~\cite{LN10,LNSW10}, relevant for foams, non-Brownian emulsions, and granular materials. These two transitions have qualitatively distinct features. 

\begin{figure}[b]
\includegraphics[width=1.\columnwidth]{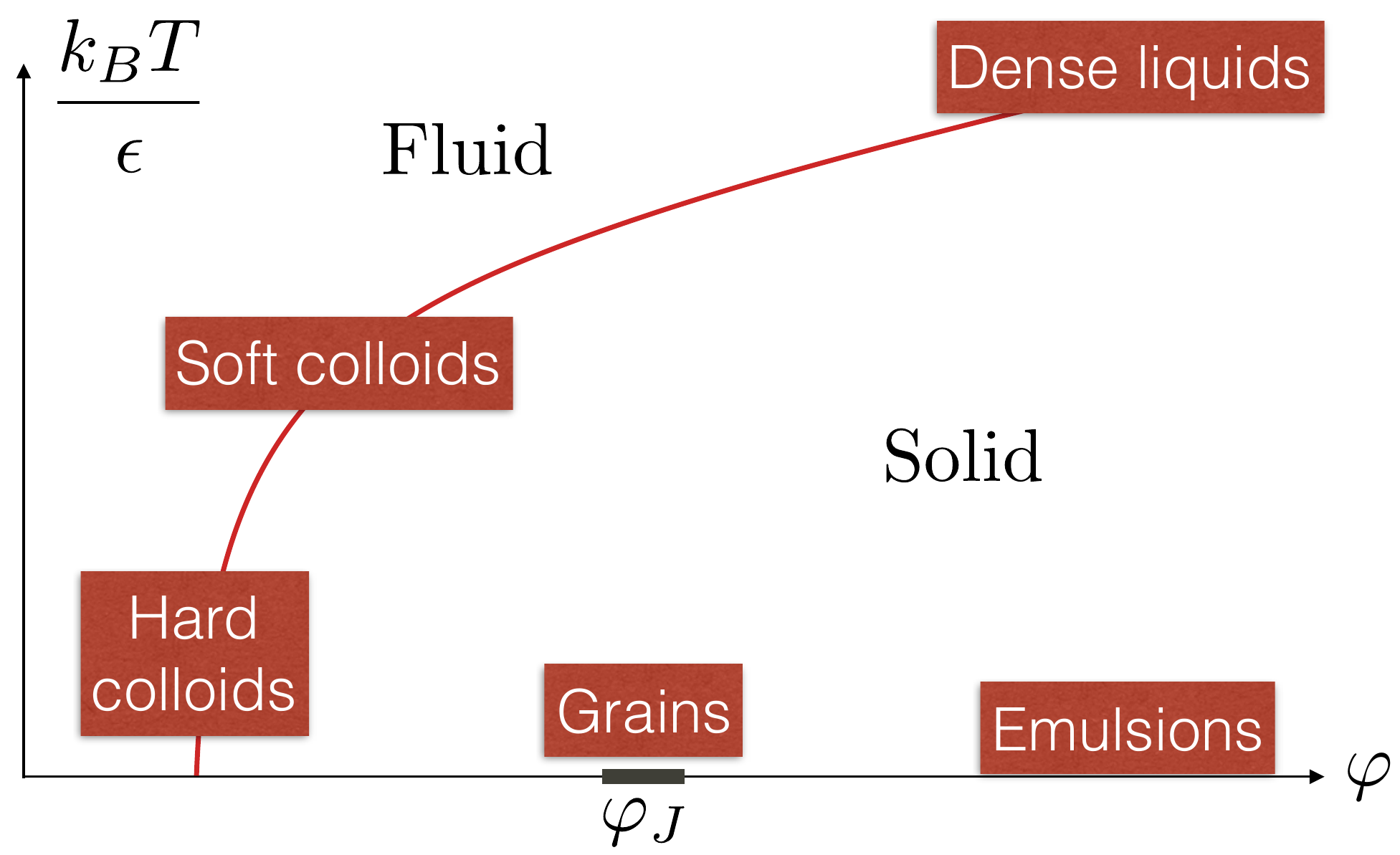}
\caption{Schematic (temperature, packing fraction) phase diagram for soft repulsive spheres and its experimental relevance. The dynamic glass transition is represented by the red line. Jamming transitions are observed in the athermal limit over a protocol-dependent range of packing fractions (grey line). Different regions of the phase diagram are relevant for a variety of amorphous materials, indicated in boxes. In this work, we explore in which conditions these amorphous materials become marginally stable.}
\label{fig:1}
\end{figure}

Models of soft repulsive spheres faithfully capture this diversity~\cite{BW09,IBS12}, as shown in Fig.~\ref{fig:1}. The relevant adimensional control parameters are the packing fraction, $\phi$, and the ratio of thermal agitation, $k_B T$, to the interaction strength between particles, $\epsilon$. A dense assembly of soft particles transforms into a glass when thermal fluctuations decrease. Glasses can also be obtained by compression at constant temperature, and in particular the limit $\epsilon \to \infty$ at constant $T$ corresponds to compression of colloidal hard spheres. At large density and temperature, the particles constantly overlap and the system behaves identically to glass-forming liquids. Intermediate densities and temperatures describe the glass transition of soft colloids. Jamming transitions are observed in the athermal regime $k_B T/\epsilon \to 0$ relevant for granular materials, foams and non-Brownian emulsions. Because this occurs deep inside the glassy phase at $T=0$, jamming transitions are protocol-dependent and occur over a continuous range of packing fractions $\f_J$~\cite{MKK08,CBS09,PZ10,OKIM12}. 

The phase diagram in Fig.~\ref{fig:1} organizes the physics of a broad variety of materials by describing how fluids lose their ability to flow, but incorrectly suggests that the solid phase has similar properties across a broad range of physical conditions. 
In fact, while ordinary glasses formed by cooling dense liquids behave roughly as crystalline solids with a high density of defects~\cite{ZP71,CP88},
glasses formed by compressing granular materials or non-Brownian emulsions across their jamming transition display unique properties distinct from ordinary solids~\cite{LNSW10,LN10}. 
For example, they may respond to weak stresses with very large deformations,
and their low-frequency excitations are very different from phonons~\cite{SLN05,LDDW14,CCPPZ16}. These properties were theoretically explained by invoking {\it marginal stability}~\cite{WNW05,WSNW05}: because these glasses are formed by zero-temperature compression across a rigidity transition, they have barely enough contacts to be mechanically stable. From this observation, several anomalous properties of athermal glasses in the vicinity of jamming can be understood~\cite{MW15}.

Theoretical calculations in the framework of the mean-field theory of the glass transition~\cite{KW87,KT87,MP09,PZ10,WL12} have 
confirmed these ideas, and suggested in addition the existence of two distinct types of amorphous solids separated by a 
sharp phase transition~\cite{KPUZ13,CKPUZ16}. 
One phase is the normal glass, and corresponds to a free energy basin that responds essentially elastically to perturbations, as any regular solid. The second is a {\it Gardner glass}~\cite{GKS85,Ga85}. The Gardner glass is marginally stable due to 
full replica symmetry breaking, as in mean-field spin glasses~\cite{MPV87}. 
Physically, marginal stability implies the existence of long-ranged correlations in the vibrational dynamics~\cite{BCJPSZ16,IBB12}, an excess of low-frequency modes~\cite{FPUZ15,FMPS18}, unusual rheological properties~\cite{BU16,FS17,JY17}, and system-spanning responses to weak, localized perturbations, manifested for instance by diverging mechanical susceptibilities~\cite{HKLP11,BU16,PRSS16}. The Gardner phase may thus provide an elegant route to understand the nature of a multitude of experimental observations of glassy excitations~\cite{KPUZ13,CKPUZ16}.
Explicit mean-field calculations for the location of marginally stable glasses were carried out for hard~\cite{KPUZ13,CKPUZ16} and soft~\cite{BU16,BU17,SBZ17} potentials, providing some insight about mean-field phase diagrams. 
Furthermore, a way to take into account fluctuations around the mean field limit, within the nucleation theory associated to the Random First Order Transition
approach, has been proposed in~\cite{LW17}.

Numerical simulations and experiments in finite dimensional systems were performed to explore these theoretical ideas, yielding contrasting results. Numerical studies of three dimensional hard sphere glasses~\cite{BCJPSZ16,JY17,tseke18,SZ18}, and numerical and experimental study of two dimensional hard disks~\cite{BCJPSZ16,SD16,LB18}, have revealed a rich vibrational dynamics, with diverging lengthscales, suggestive of a Gardner phase. On the other hand, numerically cooling soft glass-formers has only revealed sparse, localized defects~\cite{SBZ17,Bea:2017}, whereas experimental studies remain inconclusive~\cite{lunkenheimer}. 
It has also been suggested that in low dimensions localized defects could induce an apparent Gardner-like phenomenology, without an underlying sharp
phase transition~\cite{SBZ17,Moore:2017}.
Overall, this recent flurry of results suggests that distinct glassy materials may have distinct properties, depending on both their preparation and location in the phase diagram of Fig.~\ref{fig:1}, thus calling for a systematic microscopic investigation of marginally stable glassy phases. This is the central goal of the present work. 

We use a microscopic mean-field theory to study thermal soft repulsive spheres in the limit of infinite spatial dimensions to systematically investigate the physical properties and marginal stability of glasses prepared in a wide range of physical conditions, covering all regimes illustrated in Fig.~\ref{fig:1}. For a glass prepared at any given location in Fig.~\ref{fig:1}, we investigate how its properties evolve under further compression and cooling, thus providing complete phase diagrams locating simple and marginally stable glasses.
We find that all glasses may become marginally stable, but Gardner phases are more easily accessible for systems close to jamming (such as grains, foams, hard and soft colloids), and for poorly annealed glasses obtained by a fast quench. The extent of the marginally stable phase depends, in all cases, on the preparation protocol. For well-annealed glasses at intermediate packing fractions, two distinct Gardner phases are predicted. Our study extends and unifies previous analytical studies~\cite{CKPUZ16,BU17,SBZ17} and will serve as a useful theoretical guide for systematic investigations of marginal stability in finite dimensional glasses, via numerical simulations or experiments. In particular, we are currently completing a three dimensional numerical study that parallels the calculations presented here~\cite{wca3d}. 

The article is organized as follows. 
In Sec.~\ref{models}, we introduce the models studied in this work. 
In Sec.~\ref{methods}, we present the theoretical methods we use. 
In Sec.~\ref{phase}, we present the results for the phase diagrams obtained for a variety of physical conditions.
In Sec.~\ref{discussion}, we discuss our results and provide some perspectives.

\section{Models for glassy materials}

\label{models}

While we are ultimately interested in the phase diagram of dense particle systems for which the spatial dimension is $d=2$ or $d=3$, we focus on assemblies of particles embedded in an abstract, but analytically tractable, space of $d \rightarrow \infty$ dimensions. In this limit, an exact solution for the thermodynamic properties of the liquid and glass phases can be obtained~\cite{KW87,CKPUZ16}.

We study several conventional interaction potentials for glass-forming materials, that allow us to interpolate between the various physically relevant limits shown in Fig.~\ref{fig:1}. In particular, it is useful to consider the following harmonic sphere model:
\begin{equation} \label{eq:potential}
\begin{split}
v_{\text{HARM}}(r) & =\frac{\epsilon}{2}\left(1-\frac{r}{\sigma}\right)^2\theta\left(\sigma - r \right) \ ,\\
\end{split}
\end{equation}
where $r$ is the inter-particle distance, $\sigma$ the diameter of particles, $\epsilon$ the repulsion strength and $\theta(r)$ the Heaviside function. 
The harmonic sphere model was first introduced to study the jamming transition~\cite{durian}, and later studied extensively at finite temperature~\cite{BW09}.
Harmonic spheres become equivalent to hard spheres when $\epsilon \to \infty$. 

To study the large density limit relevant for dense liquids, harmonic spheres are not useful, as their extreme softness gives rise to exotic phenomena that we do not wish to discuss here. Instead, it is more relevant to analyze the Weeks-Chandler-Andersen (WCA) potential, 
\begin{equation} 
\label{eq:potential2}
v_{\text{WCA}}(r) = \epsilon \left[ 1 +  \left( \frac{\sigma}r \right)^{4 d} - 2 \left( \frac{\sigma}r \right)^{2 d} \right] \theta(\sigma - r)\ ,
\end{equation}
because it resembles the harmonic potential around the cutoff $r \sim \sigma$, but behaves as a Lennard-Jones potential at smaller inter-particle distance. Our analysis shows that the WCA model yields results qualitatively similar to the harmonic model at moderate densities, and behaves as the inverse power law (IPL) potential 
\begin{equation}
\label{eq:potentialIPL}
v_{\text{IPL}}(r)  = \epsilon (\sigma/r)^{4d}  \ ,
\end{equation}
in the large density limit. Therefore we decided to concentrate on the two models in Eqs.~(\ref{eq:potential}, \ref{eq:potentialIPL}) to report our results. Technically, the harmonic potential is easier to handle, as one can go one step further analytically than for WCA, which simplifies the numerical resolution of the equations presented below. The WCA model, on the other hand, is numerically very convenient for finite-$d$ studies, which justifies our effort to study it as well. 
While the expression of the harmonic potential is the same regardless of spatial dimension, we have extended the standard definitions of the WCA and IPL models in $d=3$ to arbitrary dimension $d$. This is done because thermodynamic stability and the existence of the thermodynamic limit,
a prerequisite for performing the theoretical development described below,
 require the potential to decay faster than $r^{-d}$ in dimension $d$~\cite{RuelleBook}.
 
We consider the thermodynamic limit for $N$ particles in a volume $V$, both going to infinity at fixed number density $\rho=N/V$.  When $d=\infty$, we can consider monodisperse particles, as crystallization is no longer the worrying issue it is in finite dimensions~\cite{SDST06,VCFC09}. 
Our adimensional control parameters are the packing fraction 
$\phi = N V_d (\sigma/2)^d / V$, defined as
the fraction of volume covered by particles of diameter $\sigma$ ($V_d$ is the volume of a $d$-dimensional unit sphere),
and the scaled temperature $T/\epsilon$ (in the following, we will take $k_B=1$).
To obtain a non-trivial phase diagram in the limit $d \rightarrow \infty$, 
the packing fraction has to be rescaled as
$\wh \phi = 2^d \phi / d$.
Note that in the case of the IPL model, the form of the interaction potential leads to a unique control parameter $\Gamma = \widehat \phi /T^{1/4}$.
We also define rescaled gaps between particles $h = d(r/\sigma-1)$, and rescaled potentials $\redv(h) $ such that  $\lim_{d \rightarrow \infty} v(r) = \redv(h)$:
\begin{equation} \label{eq:potentialinfty}
\begin{split}
 \redv_{\text{HARM}}(h) & = \frac{\epsilon}{2} h^2\theta(-h) \ , \\
    \redv_{\text{IPL}}(h) & = e^{-4 h}  \ .
\end{split}
\end{equation}
We will be particularly interested in 
mean-squared displacements (MSD) between configurations. 
In finite dimensions, they are usually defined as $\DE(X,Y) = 1/N \sum_i |\bold{x}_i - \bold{y}_i |^2$, where $X$ and $Y$ represent two configurations. Finite values for the MSD in infinite dimensions are obtained by defining $\D(X,Y) = d^2\DE(X,Y)/\sigma^2$.

In the following section, we summarize the formalism that allows us to compute the thermodynamic properties and the phase diagram for the above models. 

\section{Theoretical methods}

\label{methods}

The general strategy of our work is devised to mimic the following experimental protocol. During the gradual cooling or compression of a glass-forming liquid, the equilibrium relaxation time $\tau_{\alpha}$ of the system increases very sharply. For a given protocol, there comes a moment where the system falls out of equilibrium; this represents the experimental glass transition, at state point $(T_g,\wh\phi_g)$.  
After this moment, the system follows a `restricted' equilibrium, where the amorphous structure frozen at the glass transition is adiabatically followed at different temperature and density, $(T,\wh\phi)$. Our analytical strategy follows this protocol closely. We draw an equilibrium but dynamically arrested configuration at $(T_g,\wh\phi_g)$, 
and follow its thermodynamics when brought adiabatically to another state point $(T,\wh\phi)$ within the same glass basin.

\subsection{Glass free energy}

The state-following protocol described above is possible if the relaxation time of the initial state is extremely large~\cite{FP95,ZK10_1,ZK10_2,RUYZ15,CKPUZ16}. In infinite dimensions, the equilibrium relaxation time diverges at the dynamic glass transition $T_d(\wh \phi)$, which is of the mode-coupling type~\cite{Go09,MKZ16,CKPUZ16}. Our construction, which is briefly summarized in the following, is thus devised to follow glasses created below the dynamical transition~\cite{RUYZ15,BU16}.

Let us consider an equilibrium configuration $Y$, extracted from the Boltzmann distribution at $(T_g,\wh \phi_g)$, which falls into the dynamically arrested region $T_g < T_d(\wh \phi_g)$. To construct the thermodynamics restricted to the glass state $Y$, we consider a sub-region of phase space probed by configurations $X$ constrained to remain close to $Y$. The configuration $X$ can be at a different state point $(T,\wh \phi)$, but its mean-squared distance to $Y$ is fixed to a finite value $\D(X,Y) = \D_r $.
The free energy $f_Y$ of the glass state selected by $Y$ and brought to $(T,\wh \phi)$ can be expressed
in terms of a restricted configuration integral~\cite{RUYZ15}
\beq
\begin{split}
Z\argc{T,\wh  \phi | Y, \D_r} &= \int \dd X e^{- \b V[X]} \delta( \D_r -  \D(X,Y))~~, \\
f_Y(T,\wh  \phi | Y, \D_r) &= -\frac{T}{N} \log Z\argc{ T,\wh  \phi | Y, \D_r}~~,
\label{eq:glassfreeY}
\end{split}
\eeq
where $V(X)$ is the total potential energy of the glass $X$. The glass free energy $f_Y$ in Eq.~(\ref{eq:glassfreeY}) depends explicitly on the initial glass $Y$. In the thermodynamic limit, its typical value $f_g$ is given by averaging over all equilibrium states $Y$ 
\beq
\begin{split}
 f_g(T,\wh  \phi | T_g,\wh  \phi_g, \D_r) = & -\frac{T}{N} \int \frac{\dd Y}{Z\argc{T_g,\wh  \phi_g}} e^{- \b_g V[Y]} \\
& \times \log Z\argc{T,\wh  \phi | Y, \D_r}
\end{split}
\label{eq:glassfreegeneral}
\eeq
where $Z\argc{T_g, \wh \phi_g}$ is the standard configurational integral at $(T_g,\wh \phi_g)$. The free energy has to be computed for the parameter $\D_r$ verifying $\partial_{\D_r} f_g = 0$~\cite{RUYZ15}. Note that the density dependence of the free energy is encoded by the interaction length scale $\sigma$ of the potential, which can be changed to induce a change in packing fraction.

Performing the disorder average in Eq.~(\ref{eq:glassfreegeneral}) is challenging. Translational invariance, necessary to use saddle-point and perturbative methods, is broken by the presence of disorder. To compute the glass free-energy in Eq.~(\ref{eq:glassfreegeneral}) we use the replica method, and introduce $(s+1)$ identical replicas of the original atomic system to undertake the computation~\cite{FP95,RUYZ15,CKPUZ16}. The `master' replica represents the initial glass at $(T_g,\wh \phi_g)$, while the $s$ others `slave' replicas represent the glass at $(T,\wh \phi)$. The glass free-energy can then be expressed in terms of the MSD between the different replicas. The MSD between any slave replica and the master replica are parametrized by $\D_r$. We make the simplest assumption, called replica symmetric, and consider that all slave replicas are equivalent~\cite{RUYZ15}, at a distance $\D$ from each other. At the end of the computation, we take the analytic continuation $s \rightarrow 0$ and obtain the replica symmetric glass free energy 
\beq\label{eq:fg}
-\frac{2}{d} \b f_g = \frac{2 \D_r}\D + \log(\pi \D/d^2) + \wh \phi_g \!\! \int_{-\infty}^\infty \!\! d h\, P(h) f(h) \ ,
\eeq
defining for simplicity $\eta = \log(\wh \f / \wh \f_g)$, 
\beq
q( \D,\b;h) = \int_{-\io}^\io dy \, e^{-\b \redv(y)} 
\frac{e^{- \frac{(y-h- \D/2)^2}{2 \D}}}{\sqrt{2 \pi \D}} \ ,
 \label{eq:q}
\eeq
and 
\beq\begin{split}
P(h) &= e^h \,q( 2\D_r - \D , \b_g ; h) \ ,  \\
f(h) &= \log q(\D,\b;h-\eta) \ .
\end{split}\eeq

Compressing and decompressing a glass corresponds to $\eta > 0$ and $\eta <0$, respectively. The glass free energy should be computed with the thermodynamic values for $\D$ and $\D_r$, determined by setting to zero the derivatives of $f_g$ with respect to these parameters, which provides two implicit equations for $\del$ and $\del_r$:
\beq \label{eq:DDr}
\begin{split}
& 2\Delta_r = \Delta + \wh\f_g \D^2 \int_{-\infty}^\infty  d h\, \frac{\partial}{\partial \Delta} \left[ P(h)f(h)\right] \ , \\
&\frac{2}{\Delta}\! = -\wh \f_g  \int_{-\infty}^\infty  d h\,  f(h) \frac{\partial}{\partial \Delta_r} P(h)  \ .
\end{split}
\eeq
\subsection{Dynamic glass transition}

Our method focuses on glasses prepared at $(T_g, \wh \phi_g)$, below the dynamical transition. Our first task is thus to compute the dynamical transition line, $T_d = T_d(\wh \phi)$ for the models presented in Sec.~\ref{models}. To do so, let us consider the special case $(T_g,\wh \phi_g) =(T,\wh \phi)$ in the above construction. In that case, $\D = \D_r \equiv \D_g$ is solution of $f_g$ in Eq.~(\ref{eq:DDr}) if the glass MSD $\D_g$ verifies
\beq
\begin{split}
\frac{1}{\wh\f} &= - \D_g \!\! \int_{-\io}^\io\!\!\!\! dh \, e^{h} 
 \log q(\D_g,\b;h) \frac{\partial q(\D_g,\b;h)}{\partial \D_g} \\
& \equiv  \FF_{\b}\!(\D_g) ~.
\end{split}
 \label{eq:delta}
 \eeq
For the models considered here, the function $\FF_\b(\D)$ is positive, vanishes both for $\D \rightarrow 0$ and $\D \rightarrow \io$ and has an absolute maximum in between. This means that Eq.~(\ref{eq:delta}) has a solution at temperature $1/\b$ only if $1 / \wh \f$ is smaller or equal to the maximum of $\FF_\b$ with respect to $\D$. Glassy states at $T$ thus exist only at packing fractions higher than $\wh \f_d$, defined by
\beq \label{eq:dynamical}
1/\wh\f_d =  \max_{\D} \FF_{\b}( \D) \ .
\eeq

\begin{figure}[t]
\includegraphics[width=1.\columnwidth]{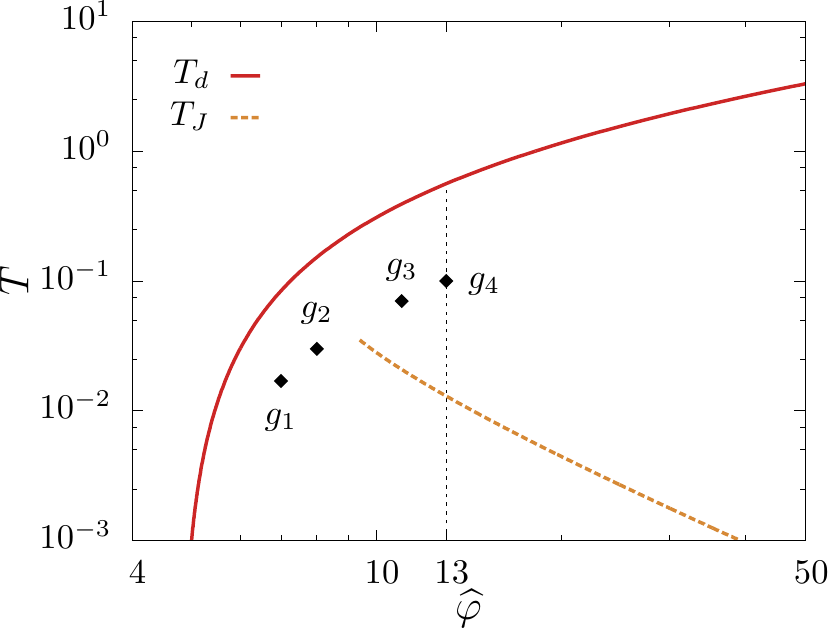}
\caption{Equilibrium mean-field phase diagram for harmonic spheres. The dynamical transition $T_d$ (red line) separates liquids that flow (above) from dynamically arrested ones (below). We select equilibrium glasses in the dynamically arrested region, for example $g_1,..., g_4$, and follow each glass adiabatically in temperature and packing fraction. The corresponding state-following phase diagrams are presented Fig. \ref{fig:5}(a-d). Glasses equilibrated above the line $T_J$ (dashed line) are jammed once minimized to $T=0$, while glasses selected below the line are unjammed at $T = 0 $. The state-following phase diagrams of glasses prepared at $\wh \f_g = 13$ (vertical dashed line) are presented in Figs.~\ref{fig:3}-\ref{fig:4}.}
\label{fig:2}
\end{figure}

We numerically solve Eq.~(\ref{eq:dynamical}) for all temperatures and find the dynamical transition line $\wh \f_d(T)$, or equivalently $T_d(\wh \f)$. 
The result is represented for the harmonic potential in Fig.~\ref{fig:2}. The line separates liquids that flow from dynamically arrested ones. The qualitative behavior of $T_d(\wh \f)$ in the WCA model is similar to that of harmonic spheres presented in Fig.~\ref{fig:2}. In both cases, the dynamical transition temperature is an increasing function of $\wh \f$, and is defined for $\wh \f > 4.8067$, which corresponds to the dynamical transition for hard spheres~\cite{PZ10}. In the large density limit, the WCA model behaves as the inverse power law potential, and the dynamical transition scales as $T_d \sim \wh \f^4$. The coefficient of proportionality is $1/\Gamma_d^4$, where $\Gamma_d = 4.304$ is given by the dynamical transition of IPL glasses.

\subsection{Adiabatically following the glass properties}

We focus on glasses prepared at $(T_g, \wh \phi_g)$ in the dynamically arrested phase. We study their thermodynamic properties when adiabatically brought to temperature and packing fraction $(T, \wh \phi)$. In particular, we compute the average potential energy per particle $\wh e_g$, given by the derivative of $f_g$ in Eq.~(\ref{eq:fg}) with respect to the inverse temperature
\beq \label{eq:energy}
 \wh e_g  = \frac{1}{d}  \frac{\partial (\b f_g)}{\partial \b}  = - \frac{ \wh\f_g}2  \int_{-\infty}^\infty  \!\!d h\,P(h) \frac{\partial}{\partial \b} f(h)  \ . 
\eeq
The energy is to be computed using the thermodynamic values for $\D,\D_r$, which solve Eqs.~(\ref{eq:DDr}). 

We employ the following strategy to numerically solve the equations, find the values of $\D$ and $\D_r$ at each state point, and consequently compute the glass potential energy. First, we compute the MSD $\D_g$ of the glass at $(T_g, \wh \phi_g)$, by numerically solving Eq.~(\ref{eq:delta}). Starting at $(T_g, \wh \phi_g)$ with the initial condition $\D = \D_r = \D_g$, we gradually change the temperature and/or packing fraction by small steps towards $(T, \wh \phi)$. At the beginning of each step, we use the values $\D,\D_r$ of the previous step as initial guesses. We then solve iteratively Eqs.~(\ref{eq:DDr}) by computing the right hand side of the equations to obtain new estimates of $\D$ and $\D_r$ until convergence is reached. We repeat this procedure until the final state $(T, \wh \phi)$ is reached.

\subsection{Gardner transition}

The glass free energy $f_g$ defined in Eq.~(\ref{eq:fg}) is derived assuming that the symmetry under permutations of replicas remains unbroken. At each state point, we must check the validity of this assumption. In practice, we check that the replica symmetric solution is a stable local minimum of the free energy. The replica symmetric solution becomes locally unstable against replica symmetry breaking when one of the eigenvalues of the stability operator of the free energy changes sign~\cite{MPV87}. This so-called replicon eigenvalue can be expressed in terms of $\D,\D_r$ as follows~\cite{RU16}
\beq
\lambda_R=1 -\frac{\wh \varphi_g}{2} \D^2  \int_{-\infty}^\infty \!\! d h \, P(h) f''(h)^2 \ .
\eeq
At each state point, the converged values for $\D,\D_r$ are used to compute the replicon eigenvalue. In the replica symmetric, or simple glass phase, the replicon is positive. The replicon might become negative upon cooling or compressing a glass, signaling its transformation to a replica-symmetry broken glass. We show that in most cases, the simple glass transforms into a marginally stable glass, characterized by full replica symmetry breaking (fullRSB). This is a Gardner transition, in analogy to a similar phase transition found at low temperature in some spin glasses~\cite{Ga85,GKS85}. 

In the marginally stable phase a complex, full replica symmetry breaking, solution should be used to derive accurately the thermodynamics of the glass~\cite{RU16}. 
Such solution is parametrized by a function $\D(x)$, for $x\in [0,1]$, associated to the distribution of mean-squared distances between states. While computing
the full function $\D(x)$ requires a rather heavy numerical procedure~\cite{RU16}, one can estimate its shape close to the transition where $\lambda_R=0$,
by a perturbative calculation~\cite{So85,FPSUZ17}. One gets
\beq
\D(x) \sim \begin{cases}
\D(\lambda) - \epsilon \dot \D(\lambda)   & x< \lambda - \epsilon\ , \\
\D(\lambda) + \dot \D(\lambda) (x-\lambda) & \lambda - \epsilon < x < \lambda + \epsilon \ , \\
\D(\lambda) + \epsilon \dot \D(\lambda)  & x>\lambda + \epsilon \ . 
\end{cases}
\eeq
Here, $\lambda$ is called the {\it breaking point} or {\it MCT parameter}.
It is related to the mean field dynamical critical exponents of the transition~\cite{CFLPRR12,FJPUZ12,KPUZ13,PR13} and, presumably, to the universality
class of the transition beyond mean field theory~\cite{PRR14}. 
At the transition point, $\epsilon \to 0$, and the constant RS solution $\D(x)=\D(\lambda)=\D$ is recovered. 
Because $\D(x)$ must be monotonically decreasing for $x\in [0,1]$, a consistent fullRSB solution
requires $\lambda\in [0,1]$ and ${\dot\D(\lambda) < 0}$.
The perturbative calculation gives~\cite{So85,FPSUZ17},
\beq
\label{eq:breaking}
\begin{split}
\lambda& =\frac{ \wh \varphi_g\int_{-\infty}^\infty dh \, P(h)  f'''(h)^2}
{\frac{4}{\D^3} + 2\wh \varphi_g\int_{-\infty}^\infty dh \, P(h)  f''(h)^3} \ , \\
\dot\D(\lambda) &= \frac{ \frac4{\D^3} + 2 \int_{-\io}^\io dh P(h) f''(h)^3 }{\frac{12 \lambda^2}{\D^4}-  \int_{-\io}^\io dh P(h) A(h)  } \ , \\
\text{with} \\
A(h) & = f''''(h)^2 - 12 \lambda f''(h) f'''(h)^2 + 6 \lambda^2 f''(h)^4 \ ,
\end{split}\eeq
which should be evaluated at the transition point. We systematically compute the value of the breaking point $\lambda$ and slope $\dot\D(\lambda)$ at the point where $\lambda_R = 0$ in order to characterize the type of symmetry breaking transition.
If $\lambda \in [0,1]$ and $\dot\D(\lambda) <0$ it is a Gardner transition.  
If instead $\lambda \in [0,1]$ but $\dot\D(\lambda) >0$, the transition is likely to be continuous towards a non-marginal 1RSB phase~\cite{FPSUZ17}.

In the following, 
we will show results for the boundary between simple and replica-symmetry broken phases (1RSB and fullRSB),
without further solving the thermodynamics of the glass inside the replica-symmetry broken phase.
Note that here we are mostly interested in the location of the marginally stable fullRSB glass phase. 

\subsection{Spinodal transition}

A glass prepared at $(T_g, \wh \phi_g)$ can also be followed upon heating ($T > T_g$), or in decompression ($\wh \f < \wh \f_g$, equivalently $\eta <0$). In that case, the glass energy becomes lower than the one of the liquid, until a spinodal transition is reached at ($T_{sp}, \wh \f_{sp}$). In practice, the spinodal transition is found when the solution for $\D,\D_r$ disappears through a bifurcation. This spinodal transition physically corresponds to the melting of the glass into the liquid. At the spinodal transition thermodynamic quantities display a square-root singularity, for instance   $\wh e_g \sim \sqrt{T_{sp} - T}$. 

Note that the replica symmetric solution also displays an unphysical spinodal transition in the region where it is unstable against fullRSB~\cite{RUYZ15,SBZ17}. 
This spinodal is unphysical because, for example, one finds that a glass might become unstable and melt upon cooling, which is physically inconsistent. The correct computation of the stability limit in the region where the replica symmetric solution is unstable should be done by solving the fullRSB
equations, which goes beyond the scope of this work.
In the phase diagrams we will show in the following, we will not draw the replica symmetric spinodal in the region where the replica solution is unstable.

\subsection{Jamming transition}

The harmonic and WCA potentials Eqs.~(\ref{eq:potential}, \ref{eq:potential2}) define a physical size for the particles. Dense assemblies of particles interacting via these two potentials will therefore have a jamming transition at $T=0$ and some packing fraction. For each studied glass, we find the location of its corresponding jamming transition point at the replica-symmetric level. To do so, we monitor the potential energy $\wh e_g$ of the glass, Eq.~(\ref{eq:energy}), down to $T=0$. Depending on its value at $T=0$, we either compress (if $\wh e_g(T = 0) = 0$) or decompress (if $\wh e_g(T = 0) > 0$) the zero-temperature packing until we reach the packing fraction $\wh \f_J$ at which the energy changes from a finite value to zero. The jamming transition of the initial glass occurs at $(T=0,\wh \f_J)$, or equivalently at $(T=0,\eta_J)$. 

We stress that the location of the jamming transition depends on the specific choice of the state point $(T_g, \wh \phi_g)$ at which the glass was prepared in the phase diagram of Fig.~\ref{fig:2}. It is useful to define an additional line $T_J(\wh \f_g)$ in the phase diagram to rationalize the results in  Sec.~\ref{phase}. This line separates glasses into two classes: if $T_g > T_J(\wh \f_g)$, the state is jammed at $T=0$ and $\wh e_g(T = 0) > 0$, while if $T_g < T_J(\wh \f_g)$, the state is unjammed at $T=0$ and $\wh e_g(T = 0) = 0$. We compute this line by taking analytically the zero-temperature limit of Eqs.~(\ref{eq:DDr}-\ref{eq:energy}), and solving them numerically for all initial equilibrium glasses. 

\begin{figure}[t]
\includegraphics[scale = 1]{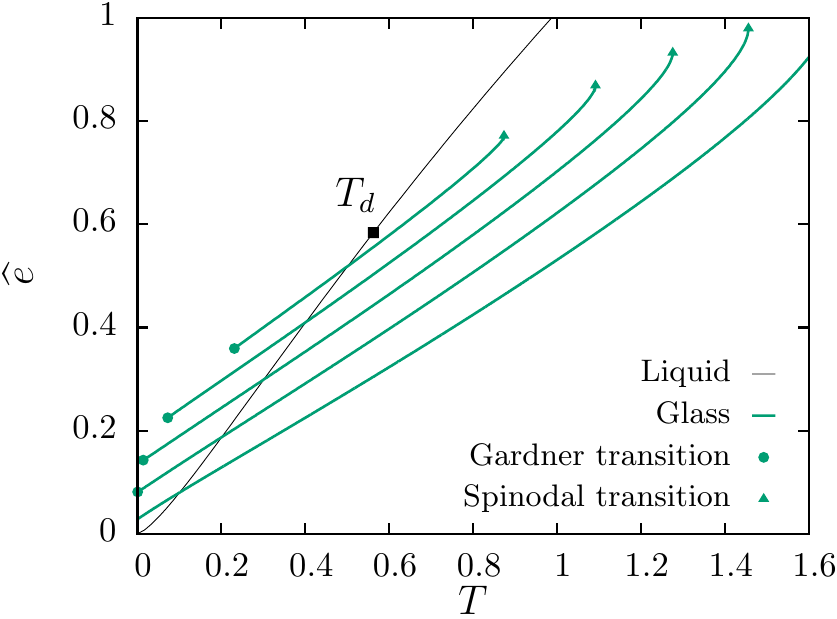}
\caption{Energy per particle $\wh e$ of the equilibrium liquid and several glasses selected at $\wh \phi_g = 13$ (vertical dashed line in Fig.~\ref{fig:2}), as a function of temperature $T$ for constant $\wh\f = \wh\f_g$. The energy of the liquid is given by the thin black line, on which lies the dynamical transition at $T_d = 0.562$ (black square). The energy of simple glasses created at $T_g < T_d$ ($T_g = 0.5$, 0.4, 0.3, 0.2, 0.1, from top to bottom) are represented by green lines. Upon cooling, these glasses may undergo a Gardner transition (bullets) to a marginally stable glassy phase, in which the equation of state must be computed solving the fullRSB equations (not shown). When heated, the glasses remain stable up to a temperature $T_{sp}$ (triangles) at which the glass melts into the liquid.}
\label{fig:3}
\end{figure}

The resulting line $T_J(\wh \f_g)$ for harmonic glasses is represented in Fig.~\ref{fig:2}. This line is qualitatively similar for WCA glasses. In both models, $T_J$ is a decreasing function of $\wh \f$: starting from better annealed glasses (lower $T_g$) shifts the jamming transition of the glass to higher packing fractions. This feature is also observed in the phase diagram of infinite dimensional hard sphere glasses. The line $T_J$ should in principle extend to lower packing fractions and reach $T_d$. This is not the case in Fig.~\ref{fig:2}, as glasses prepared in this region present an extended marginal phase at finite temperature (for example, see Fig.~\ref{fig:3}), and the replica symmetric solution is lost before reaching $T = 0$. Using a fullRSB solution, we would find that this line extends smoothly at lower densities until hitting the dynamical transition line.

\section{State-following phase diagrams}
\label{phase}

We now present how glasses prepared in a wide range of conditions evolve when subject to cooling/heating or compression/decompression, or a combination of both. We are particularly interested in finding the boundaries of the marginally stable phase. In Secs.~\ref{cool_harmonic} to \ref{jamming_gardner}, we present results for the harmonic sphere model. Equilibrium glasses at $(T_g,\wh \f_g)$ are chosen in the region delimited by the dynamical transition in Fig.~\ref{fig:2}. For each initial glass, we construct a two-dimensional state-following phase diagram, presented in terms of $T$ and $\eta$. 
Results for the inverse power law are presented in Sec.~\ref{dense}: in this case, the representation is easier because there is a single control parameter
$\Gamma = \widehat \phi /T^{1/4}$. As stated above, the WCA potential would yield results similar to harmonic spheres for densities close to jamming, but similar to the inverse power law potential at large densities. We will present selected state-following results that highlight the main features of these phase diagrams, and propose a representation which summarizes the most important findings (see Fig.~\ref{fig:6}). 

\subsection{Cooling and heating glasses}\label{cool_harmonic}

We first focus on heating and cooling glasses prepared at an intermediate packing fraction, $\wh \phi_g = 13$, and several temperatures $T_g$. These equilibrium initial states are selected along the vertical dashed line displayed in the phase diagram in Fig.~\ref{fig:2}. 

We present the results in terms of potential energy per particle $\wh e$ as a function of temperature in Fig.~\ref{fig:3}, with the density being kept constant
at its original value, $\wh\f=\wh\f_g$. 
The energy of the equilibrium liquid is computed, along with the dynamical transition at temperature $T_d= 0.562$. We select glasses within a large range of glass stabilities, prepared at $T_g = 0.5$, 0.4, 0.3, 0.2, 0.1. We then follow their energy as a function of temperature, and report the corresponding glass equations of state
in Fig.~\ref{fig:3} (colored lines). 
Note that all the glasses presented in Fig.~\ref{fig:3} have a strictly positive potential energy at zero temperature. 
Indeed, they have all been prepared at temperatures $T_g$ higher than $T_J (\wh \f_g =13) = 0.013$.

Upon cooling, the simple glass may destabilize when the replicon vanishes. 
The slope $\dot\D(\lambda)$ is formally positive for $T_g>T_g^\dag\simeq 0.524$, indicating that
glasses prepared near the dynamical transition $T_g^{\dag } < T_g < T_d$ undergo a continuous one step replica-symmetry breaking (1RSB) transition
towards a non-marginal phase.
We find instead that for glasses prepared at  $T_g < T_g^{\dag } $, such as those presented in Fig.~\ref{fig:3}, 
the slope $\dot\D(\lambda)$ is negative at the transition.
The simple glass thus transforms into a marginally stable glass at a Gardner transition, reported with bullets in Fig.~\ref{fig:3}. The breaking point $\lambda$ computed with Eq.~(\ref{eq:breaking}) at the Gardner transition equals $\lambda = 0.315$, 0.159, 0.068, 0.01 for $T_g = 0.5$, 0.4, 0.3, 0.2, respectively. 
Note that $\lambda\to 0$ when the Gardner transition temperature $T_G\to 0$, while $\dot\D(\lambda)\to-\io$ when
$T_g\to T_g^\dag$ from below.
We observe that the glass is marginally stable over a large temperature range when prepared at higher $T_g$. The extent of the marginally stable region diminishes for better annealed glasses (decreasing $T_g$). The Gardner transition temperature $T_G$ of a given glass decreases with decreasing $T_g$, so that better annealed glasses remain stable down to lower temperatures. For the most stable glass reported in Fig.~\ref{fig:3}, prepared at $T_g = 0.1$, the glass remains stable down to zero temperature, and no marginally stable phase is observed when cooling. When glasses are instead heated, their energy follows the glass equation of state and remains smaller than the energy of the liquid up to the spinodal transition $T_{sp}$ at which the glass melts into the liquid. The temperature range over which the glass remains stable increases when the glass transition temperature $T_g$ decreases, which is the experimental hallmark of increasing glass stability~\cite{MDE12,MDE17,FB17}.

Overall, increasing the degree of annealing of the glass extends the region of stability of the simple glass phase, pushing the marginal phase to lower (possibly vanishing) temperatures and the spinodal transition to higher temperatures.
\begin{figure}[t]
\includegraphics[scale = 1]{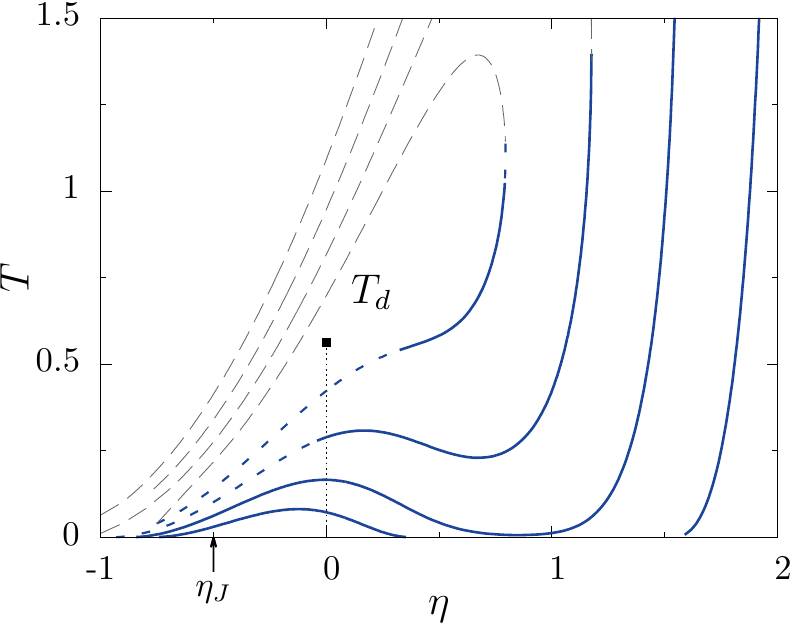}
\caption{Glasses prepared at $\wh \f_g = 13$,  $T_g  = 0.55$, 0.52, 0.47, 0.4 (top to bottom) are followed in temperature $T$ and packing fraction $\wh \f$, expressed as $\eta=\log(\wh \f /\wh \f_g)$. The dynamical transition is indicated with a square.
For each glass, we show the limits of stability of the simple glass. The simple glass loses it stability and melts at the glass spinodal (dashed grey). The simple glass also destabilizes at the Gardner transition (solid blue), or at a continuous transition towards a 1RSB phase (dashed blue). Below the Gardner transition line, the glass is marginally stable. 
}
\label{fig:4}
\end{figure}
\subsection{Temperature-density glass phase diagram}
\label{T_g}

The results of thermal quenches shown in Fig.~\ref{fig:3} give only a partial view of the state-following phase diagrams, because density is not varied. We now study how the marginally stable phase extends both in temperature and packing fraction. Specifically, we present how glass stability modifies the extent and nature of the marginally stable phase.
We compute  state-following phase diagrams for glasses prepared at $\wh \f_g = 13$ and different annealing, $T_g = 0.55$, 0.52, 0.47, 0.4. For each glass, we compute the Gardner transition line $T_G(\eta)$ at which the glass becomes marginally stable, and we report it as a blue line in Fig.~\ref{fig:4}. As in Fig.~\ref{fig:3}, less annealed glasses first transform to a 1RSB glass, which we indicate with a blue dashed line. We expect the 1RSB glass to transform to a marginally stable fullRSB glass at lower temperature.
For each $T_g$, we can also compute the replica symmetric spinodal where the glass melts into the liquid, also reported in Fig.~\ref{fig:4} as a grey dashed line. 
For a given $T_g$, the region delimited by the solid and dashed lines defines the simple (replica symmetric) glass region.
At temperatures below the blue line, the marginal (fullRSB) glass phase exists. This phase is delimited by the blue line, and by fullRSB spinodal lines
that continue the grey line at lower temperatures; unfortunately, these lines can only be computed by solving the fullRSB equations, which goes beyond the scope
of this work. 
We thus interrupt the spinodal grey line when it crosses the Gardner line, but the reader should keep in mind that this line should be continued at lower temperature
to properly delimit the marginal glass phase.
\begin{figure*}[t]
\begin{center}
\includegraphics[scale = 1]{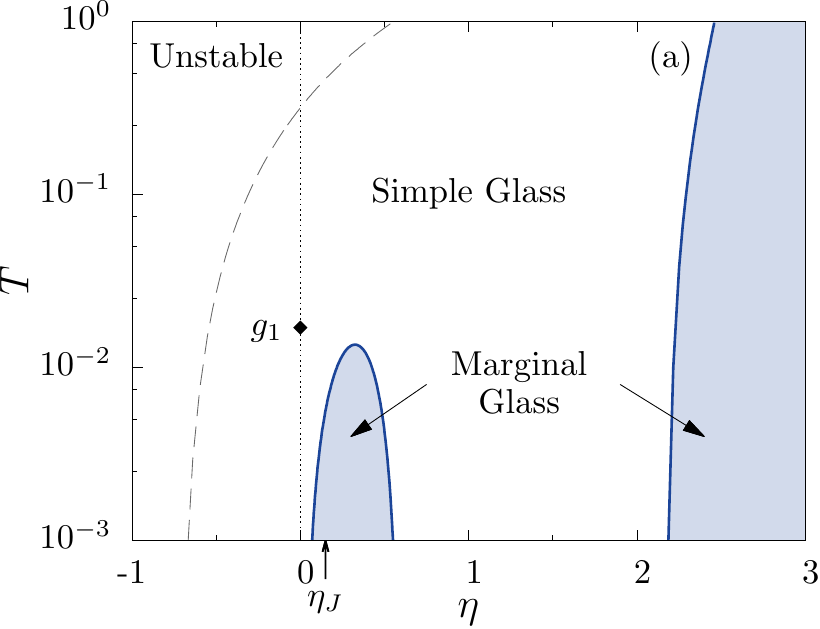}
\includegraphics[scale = 1]{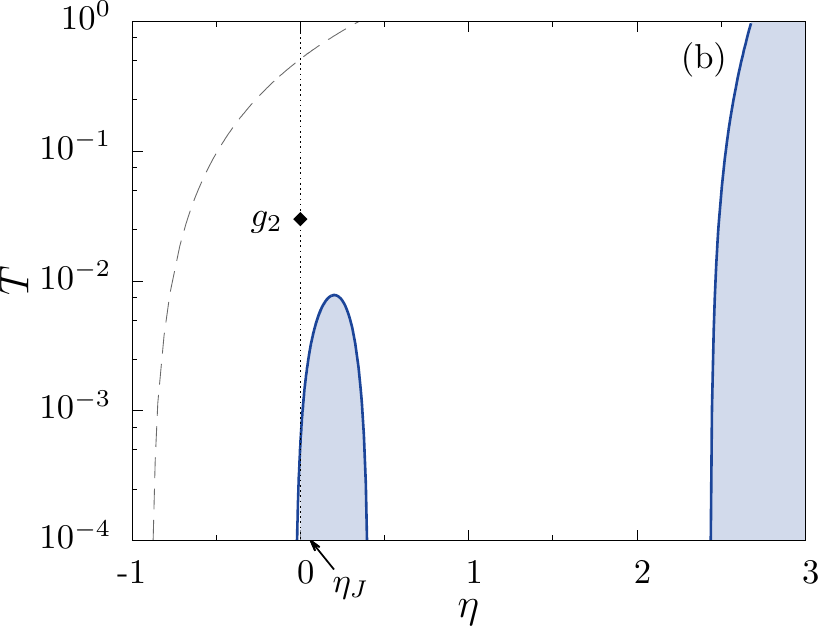}
\includegraphics[scale = 1]{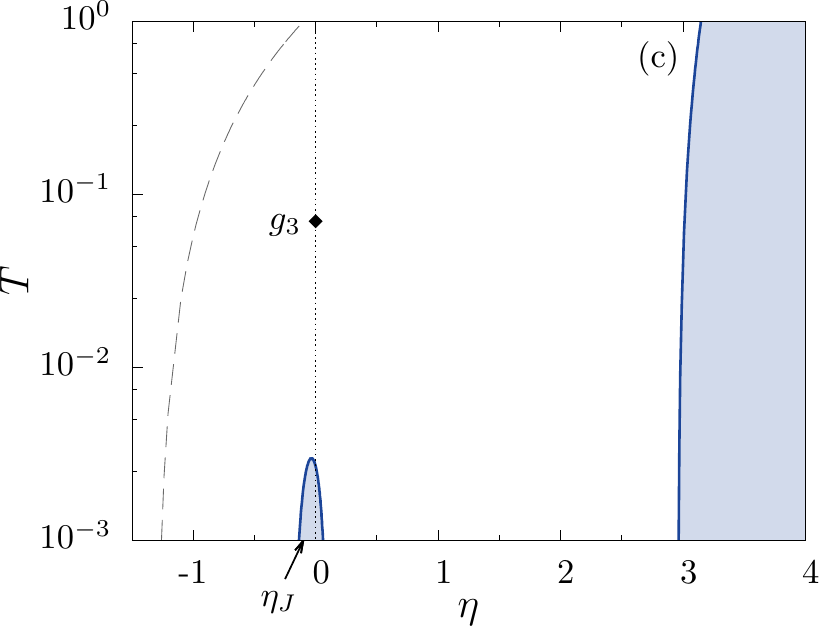}
\includegraphics[scale = 1]{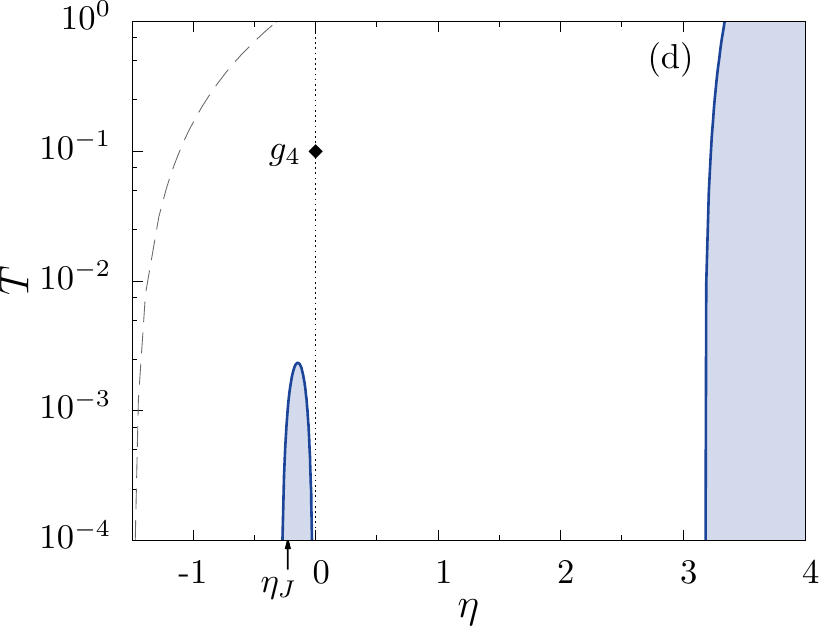}
\caption{Mean-field state-following phase diagrams for four different starting harmonic glasses: (a) $g_1$, (b) $g_2$, (c) $g_3$, (d) $g_4$, whose location in the equilibrium phase diagram is shown Fig.~\ref{fig:1}. The region of stability of the simple glass phase is delimited by the spinodal (dashed grey line) and Gardner transition line (full blue line). Above the spinodal, the glass melts into the liquid. Below the Gardner transition line, the glass is marginally stable (shaded blue region). The jamming transition of the glasses takes place at $T= 0$, and $\eta_J$ indicated by an arrow.}
\label{fig:5}
\end{center}
\end{figure*}
Glasses prepared exactly at the dynamical transition, $T_g = T_d$, are unstable towards RSB everywhere in the glassy phase. We see in Fig.~\ref{fig:4} that the unstable phase of glasses prepared slightly below $T_d$ (top curve corresponding to $T_g=0.55$) still extends over a large region of the state-following phase diagram. As the glass preparation temperature decreases, the unstable phase becomes everywhere marginally stable, and its extension diminishes. This observation is consistent with the results of the previous subsection, but Fig.~\ref{fig:4} reveals a new, more subtle, phenomenon. The shape of the Gardner transition line evolves qualitatively as $T_g$ decreases. While the Gardner transition line $T_G(\eta)$ of the less stable glasses (top curves in Fig.~\ref{fig:4}) increases monotonically with $\eta$, it becomes non-monotonic for lower $T_g$. 
For very well annealed glasses, such as $T_g = 0.4$, the line even forms two disconnected regions. The marginal phase then comprises a `dome' around the jamming transition occurring at $\eta_J = \log(\wh \f_J / \wh \f_g)$, and a second region located at high compression $\eta$, as also observed in~\cite{BU17}. 
The Gardner transition line which defines the latter region is qualitatively similar to the one found for the less stable glasses, but it is shifted to much higher packing fractions.

We argue that these two distinct marginally stable phases have a different character. The Gardner phase surrounding the jamming transition is similar to the one found by compressing hard sphere glasses. The presence of a Gardner phase is crucial for an accurate mean-field description of jamming.
The marginally stable phase at high compression appears as a remanent of the marginality which exists near the dynamical transition. It is always present, and increasing the glass stability only shifts that phase to higher density.
Finally, these two distinct phases would also be present for the WCA pair potential over a range of intermediate densities, because WCA particles and harmonic spheres have the same behavior in this regime. However, WCA particles behave qualitatively differently at large densities, as described below in Sec.~\ref{dense} where the inverse power law potential is analyzed. 

\subsection{Interplay between jamming and Gardner phase}
\label{jamming_gardner}

We have studied the state-following phase diagrams of many initial glasses prepared in a variety of conditions $(T_g, \wh \phi_g)$. We find that the phenomenon described in the previous subsection is generically observed for glasses prepared in all regions of the glass phase. 
For well-annealed glasses, the marginally stable phase always splits into two distinct regions. We focus on four representative well-annealed glasses $g_1,...,g_4$, prepared at state points marked by black squares in Fig.~\ref{fig:2}. These glasses are stable enough that the Gardner phase is separated into two distinct regions.

We present in Fig.~\ref{fig:5}(a-d) the state-following phase diagram for each initial glass $g_1,...,g_4$. 
 We first determine the location of the jamming transition $(T=0,\wh \f_J)$ for each initial glass. The value $\eta_J = \log(\wh \f_J/ \wh \f_g)$ is indicated in Fig.~\ref{fig:5}(a-d). 
We then focus on the limit of stability of the simple glass phase. 
For all four glasses, we draw the corresponding Gardner transition lines separating the two types of glasses, which
 separates into a dome around jamming and a marginal phase at high compression.
We have checked that the simple glass always destabilizes to a marginally stable (fullRSB) glass, as the slope $\dot\Delta(\lambda)$ is always negative. 
The parameter $\lambda$ is finite at the left end of the dome (corresponding to the hard sphere Gardner transition~\cite{RU16}),
and decreases along the dome to reach $\lambda=0$ at its right end, corresponding to a zero-temperature soft sphere Gardner transition. It then increases
again from $\lambda=0$ at zero temperature, along the higher-density Gardner transition line.
The difference between the four diagrams is the relative location of all these elements.

The glasses $g_1$ and $g_2$ are prepared below the line $T_J$. Their jamming transition is therefore found by compressing the glass $(\eta_J>0)$ at $T=0$. In addition, $|\eta_J^{g_2}|<|\eta_J^{g_1}|$ because $g_2$ is prepared closer to the line $T_J$ in Fig.~\ref{fig:2}. Glasses $g_3$ and $g_4$ are prepared above $T_J$, and their jamming transition takes place when decompressing them $(\eta_J<0)$ at $T = 0$. Moreover, $|\eta_J^{g_4}| > |\eta_J^{g_3}|$ because $g_3$ is prepared closer to $T_J$ in Fig.~\ref{fig:2}. 

For the glass $g_1$, the dome surrounding jamming only appears for $\eta>0$, and this glass does not undergo a Gardner transition as it is cooled down to zero temperature at constant density. By contrast, the denser glass $g_2$ is located above the dome of marginality, and that glass can undergo a Gardner transition simply by cooling. A similar qualitative difference is observed for the glasses $g_3$ and $g_4$, both prepared above the $T_J$. The glass $g_3$ will become marginal if cooled at constant packing fraction, while the glass $g_4$ will remain stable down to its ground state. Despite these differences, all these glasses can nevertheless become marginal by a combination of cooling and compression/decompression over a broad range of state points. Finally, all these glasses also become marginal when compressed to large packing fractions far above jamming. 

\begin{figure}
\includegraphics[scale = 1]{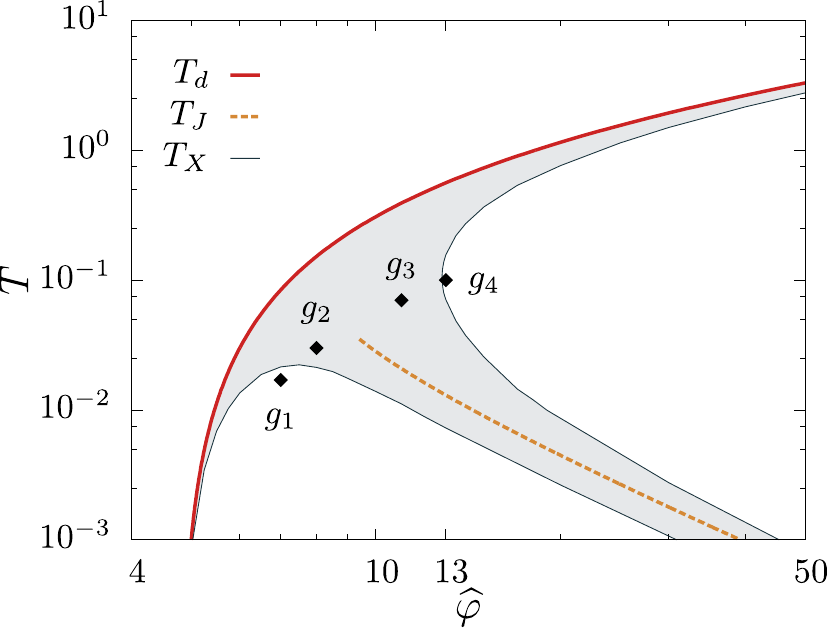}
\caption{Equilibrium mean-field phase diagram of harmonic spheres, as in Fig.~\ref{fig:2}. We add the transition lines $T_X$ (thin lines) which delimit the glasses that become 
unstable upon cooling (shaded region), such as $g_2,$ and $g_3$, or not, as for $g_1$ and $g_4$.
}
\label{fig:6}
\end{figure}

The phase diagrams found in Fig.~\ref{fig:5} suggest the existence of two types of behaviors. Some glasses undergo a Gardner transition as they are cooled, while some glasses do not. This distinction depends both on the initial temperature $T_g$ of the glass, and on its initial density $\wh\f_g$. 
To distinguish between these two types of glasses, we define a line $T_X(\wh\f_g)$ which delimits in the $(T_g,\wh\f_g)$ phase diagram. Our results for $T_X$ are reported in Fig.~\ref{fig:6}.
Glasses prepared in the shaded part of this phase diagram, like $g_2$ and $g_3$ undergo a Gardner transition to a marginally stable phase upon cooling at constant density. The other glasses, like $g_1$ and $g_4$ do not and remain stable glasses down to $T = 0$. The corresponding phase diagram presented in Fig.~\ref{fig:6} is rather complex, exhibiting non-monotonic re-entrant lines $T_X$. The mean-field phase diagram of soft repulsive spheres is therefore not a trivial extension of the one of  hard spheres. Figure~\ref{fig:6} shows that a Gardner phase is relevant for hard sphere glasses, for soft particles prepared not too far from either the dynamical transition $T_d$ and the temperature $T_J$, which suggests two distinct possible physical origins for the Gardner phase.  

\subsection{Dense liquid regime}\label{dense}

We now focus on the dense liquid regime modeled by the IPL potential. This also corresponds to the large density limit of the WCA model, where only the repulsive part of the Lennard-Jones interaction is physically relevant.
We follow the strategy and representation adopted in Sec.~\ref{T_g} for the harmonic spheres. 

The thermodynamic state of IPL glasses only depends on the combination $\Gamma = \widehat \phi /T^{1/4}$. The complete phase diagram for the IPL model can therefore be completely understood by fixing for instance the packing fraction and changing the temperature of the glass. For convenience, we choose $\wh \f_g = 4.304$, for which the dynamical transition takes place at $T_d = 1$. We consider glasses with different stabilities, prepared at $T_g < T_d$. Despite the one-dimensional nature of the phase diagram, we show results for IPL glasses using the same representation as for harmonic spheres, using both $T$ and $\eta$, to allow for a more direct comparison of the two types of models. By definition, all lines in this diagram exactly obey the relation $T \propto e^{4 \eta}$. 

\begin{figure}
\includegraphics[scale = 1]{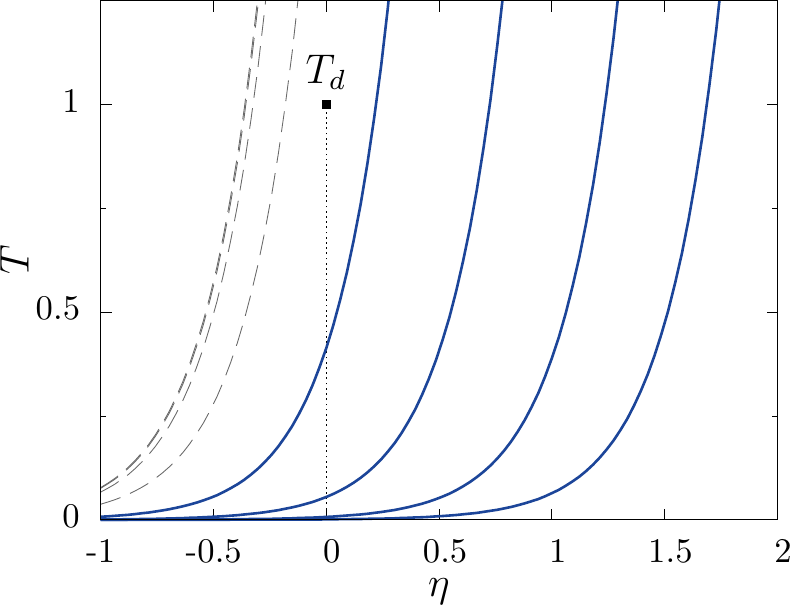}
\caption{Dense liquid regime analyzed using the IPL potential. Glasses prepared at $\wh \f_g= 4.304$ ($T_d = 1$), for various $T_g  = 0.9$, 0.8, 0.7, 0.6 are followed in the $(T, \eta)$ plane. For each glass, we represent the state-following Gardner transition line $T_G(\eta)$ (full lines) and the spinodal (dashed lines). 
All lines obey $T \sim e^{4 \eta}$. The marginal phase shifts to large density and lower temperatures as $T_g$ decreases, and disappears altogether for $T_g < 0.567$.
}
\label{fig:7}
\end{figure}

We find that glasses prepared at $T_g < T_g^{\dag} \simeq 0.92$ transforms into a marginally stable glass when cooled. Instead, glasses prepared in the range  $ T_g^{\dag} < T_g < T_d$ first transform into a 1RSB glass. As for harmonic spheres, the slope $\dot\Delta(\lambda)$ is negative for $T_g < T_g^\dag$, diverges upon approaching
$T_g^\dag$ from below, and is formally positive above it.
 The Gardner transition lines for glasses prepared at $T_g = 0.9$, 0.8, 0.7, and 0.6 are presented in Fig.~\ref{fig:7}. They have the form $T_G(\eta) = T_G(\eta=0) e^{4 \eta}$, where $T_G(\eta=0)$ is the Gardner transition temperature obtained for a simple cooling of the glass. The breaking point $\lambda$ at the Gardner transition is equal to $\lambda = 0.407$, 0.283, 0.168, 0.042 for 
  $T_g = 0.9$, 0.8, 0.7, and 0.6 respectively. As for harmonic spheres, $\lambda\to0$ when $T_G$ vanishes.
  The marginally stable phase is pushed to larger densities and lower temperatures (in fact, to larger $\Gamma$) as the glass stability increases. In this model, however, particles do not possess a physical size (the potential has no cutoff at a finite distance), and hence the jamming transition cannot be observed. As a consequence, the `domes' of marginal stability found around the jamming transition in Figs.~\ref{fig:4}-\ref{fig:5} for harmonic spheres are absent for the IPL model. 
  The behavior of the Gardner transition lines at high $\eta$ with decreasing $T_g$ is similar in the IPL and WCA models.  
  The WCA potential instead behaves as harmonic spheres near jamming and is thus characterized by domes around jamming.  

In this dense liquid regime, glasses prepared at $T_g <  0.567$ remain stable down to their ground state at $T=0$, as reported before~\cite{SBZ17}. The most stable glass for which we report the Gardner transition line in Fig.~\ref{fig:7} is $T_g = 0.6$. Below this value, glasses remain stable in the entire phase diagram and never undergo a transition to a marginally stable phase, even at arbitrarily large compressions. This is consistent with the high density/temperature limit found in the harmonic phase diagram Fig.~\ref{fig:6}, where only glasses prepared in the vicinity of the dynamical transition become marginally stable upon cooling (shaded region). However, harmonic spheres are qualitatively distinct from both WCA and IPL potential regarding compression of very stable glasses: whereas harmonic spheres always reach marginal states upon compression at constant temperature, very stable WCA and IPL glasses do not.  Note also that for harmonic spheres, the Gardner and spinodal lines meet at high density, so that the glass always melts upon large enough compression, which is not
the case of the WCA and IPL models.

\section{Discussion and perspectives}
\label{discussion}

In this work,
we have obtained the complete mean-field phase diagrams of several glass-forming models. In particular, we provided detailed information regarding the location of the marginally stable glass phases for a variety of pair interactions and physical conditions, extensively exploring physical regimes relevant to granular materials, foams, emulsions, hard and soft colloids, and molecular glasses. We find that all types of glasses may become marginally stable upon cooling or compression, but the extent of marginal phases strongly depends on the preparation protocol and the chosen model. We find that increasing the glass stability systematically reduces the extent of marginality. For well-annealed glasses, we find that marginality emerges in two distinct regions, either around the jamming transition or at high compression. Our results suggest that marginal phases should be easily observable for colloidal and non-Brownian particles near jamming, or poorly annealed glasses.

Our study unifies previous results on marginal stability in mean-field models~\cite{KPUZ13,CKPUZ16,BU17,SBZ17}. Already in mean-field theory, marginal stability emerges under distinct physical conditions in different microscopic models. This provides a way to reconcile apparently contradictory numerical and experimental studies aimed at detecting Gardner phases in finite dimensional glasses, where its existence is still debated~\cite{BU15,CY17}. In particular, the evidence for marginally stable phases reported for $2d$ and $3d$ hard spheres glasses under compression contrasts with its absence in $2d$ and $3d$ numerical models of dense liquids upon cooling. Our analysis shows that already at the mean-field level these two types of systems behave differently. 
In addition, while the critical properties around the jamming transition remain unchanged from $d = \infty$ down to $d = 2$~\cite{GLN12,CCPZ15}, the nature of the mean-field dynamical transition is highly altered by finite dimensional fluctuations~\cite{KTW89}. For instance, our results predict that highly compressed dense liquids should be marginally stable (see also~\cite{LW17}), a protocol that was never tested in finite dimensional studies.

Our results will be useful to guide future numerical simulations and experiments aimed at detecting marginally stable phases in finite dimensional glasses. We find that mean-field Gardner phases are not restricted to exist in the immediate vicinity of jamming, and could be more broadly relevant to a wide class of materials. We are currently numerically investigating, along the lines of this theoretical work, 
the evolution of the Gardner transition while continuously interpolating between regimes relevant to dense hard sphere glasses and dense liquids, 
using a WCA potential~\cite{wca3d}. 

Our results open a number of additional perspectives for future work. One finding is that soft sphere glasses can undergo a zero-temperature
Gardner transition, as reported in Fig.~\ref{fig:5}. A convenient protocol to observe this transition is suggested in Fig.~\ref{fig:5}d for the glass $g_4$. It can be quenched at $T=0$, where it is jammed and in the simple glass phase. It is therefore a stable harmonic energy minimum. Under decompression at $T=0$, this state undergoes a Gardner
transition before unjamming. The signature of this zero-temperature Gardner transition, if it exists in $2d$ or $3d$, would be particularly dramatic: the Hessian would develop
delocalized soft modes~\cite{FPUZ15}, and the system would start responding by intermittent avalanches~\cite{FS17} to an applied strain. A divergent correlation
length would also develop in the contact network~\cite{HLN17}. The absence of thermal fluctuations should make the study of this transition much easier than
in the thermal case.

While the nature of the mean-field Gardner transition is certainly affected in finite dimensions~\cite{BU15}, the existence of extended marginally stable phases should give rise to interesting  new physics in structural glasses. As happens in spin glasses, even if the Gardner phase transition is avoided in physical dimensions~\cite{Moore:2017}, it may still be the case that interesting physical phenomena, such as aging and non-linear dynamics, remain relevant to describe the behavior of structural glasses.

\acknowledgments

We thank G. Biroli and P. Urbani for useful exchanges. FZ and LB thank the ICTS, Bangalore for kind hospitality during the last stages of writing the paper. 
This work was supported by a grant from the Simons Foundation (\#454933, L. Berthier, \#454955, F. Zamponi).
This project has received funding from the European Research Council (ERC) under the European Union's Horizon 2020 research and innovation program (grant agreement 723955 - GlassUniversality).

\bibliographystyle{mioaps}
\bibliography{HS}

\end{document}